\documentclass[%
 reprint, 
nofootinbib,
nobibnotes,
 amsmath,amssymb,
 aps, physrev,
]{revtex4-2}

\usepackage{graphicx}% Include figure files
\usepackage{dcolumn}% Align table columns on decimal point
\usepackage{bm}% bold math
\usepackage{subfigure}
\usepackage{xcolor}
\usepackage{booktabs}
\begin{document}

%\preprint{APS/123-QED}

\title{\textbf{The metastability of lipid vesicle shapes in uniaxial extensional flow} 
}% 

\author{M.A. Shishkin} %\orcidlink{0000-0002-1243-5285}}

\affiliation{Landau Institute for Theoretical Physics of the RAS, Moscow Region, Chernogolovka 142432, Russia}%Lines break automatically or can be forced with \\
 \affiliation{HSE University, Moscow 101000, Russia}
 \email{Contact author: max.shishckin2011@gmail.com}
\author{E.S. Pikina}% \orcidlink{0000-0003-1992-4468}}%

\affiliation{Landau Institute for Theoretical Physics of the RAS, Moscow Region, Chernogolovka 142432, Russia}
\affiliation{Oil and Gas Research Institute, RAS, 119917, Gubkina 3, Moscow, Russia}

\date{\today}% It is always \today, today,

\begin{abstract}
	In this work, we investigate the elastic properties of deflated vesicles and their shape dynamics in uniaxial extensional flow. By analyzing the Helfrich bending energy and viscous flow stresses in the limit of highly elongated shapes, we demonstrate that all stable vesicle configurations are metastable. For vesicles with small reduced volume, we identify the type of bifurcation at which the stationary state is lost, leading to unbounded vesicle elongation in time. We show that the stationary vesicle length remains finite at the critical extension rate. The critical behavior of the stationary vesicle length and of the growth rates of small perturbations is obtained analytically and confirmed by direct numerical computations. The beginning stage of the unbounded elongation dynamics is simulated numerically, in agreement with the analytical predictions.
\end{abstract}

\maketitle

%\tableofcontents

\section*{Introduction}

Vesicles are volumes of fluid enclosed by a molecularly thin lipid bilayer.
These fluid containers play a critical role in countless biological processes, including the storage and transport of nutrients \cite{alberts2015essential},
cellular signaling \cite{settembre2013signals}, and exocytosis \cite{raposo2013extracellular}.
In recent years, GUVs (giant unilamellar vesicles) have become a canonical model system for the much more complex, protein-rich membranes of living cells \cite{Dimova2020, noguchi2005shape, fenz2012giant}.
Moreover, lipid vesicles have found widespread application in various technological fields \cite{jesorka2008liposomes, Dimova2020}, including their use as biocompatible drug-delivery capsules \cite{huang2004acoustically, muzzalupo2015niosomal}.
In these applications, a fundamental issue is the determination of vesicle shape dynamics in response to external influences.

The main physical properties of lipid bilayers have been established in numerous experimental studies \cite{Seifert1997, Dimova2020}.
First, the bilayer behaves as a weakly compressible liquid film, because there is an equilibrium surface concentration of molecules.
In addition, due to the low permeability of the bilayer to most molecules in the surrounding fluid, the vesicle volume is fixed by the condition of zero osmotic pressure difference \cite{Helfrich1973}.
The leading contribution to the surface free energy density arising from shape deformations of a vesicle can be expressed in terms of local radii of curvature and is known as the Helfrich energy in the physics literature \cite{Helfrich1973}, or equivalently as the Willmore energy in mathematics \cite{willmore1965note}.
The equilibrium shape of a vesicle in the absence of hydrodynamic motion is determined by minimizing the bending energy at fixed surface area $S$ and enclosed volume $V$.
Due to the scale invariance of the Helfrich energy, the equilibrium shape is fully characterized by the dimensionless reduced volume $\mathcal{V}=V/V_S \le 1$, where ${V_S = {4}/{3}\pi (S/(4\pi))^{3/2}}$ is the volume of a sphere with surface area $S$ \cite{Seifert1991, Lipowsky1991, Seifert1995}.
The coexistence of two conserved quantities, area and volume, leads to nonspherical equilibrium shapes, such as prolate, oblate, discocyte, and stomatocyte forms \cite{Seifert1997}.

In recent years, the non-equilibrium dynamics of vesicles in external flows has become a subject of intensive research \cite{Abreu2014, Dimova2020},
relevant to both microfluidic devices and biological systems.
Most studies have focused on vesicle dynamics in steady shear flows, revealing three distinct dynamic regimes: tumbling, trembling, and tank-treading.
Theoretical models \cite{seifert1999fluid, misbah2006vacillating, Lebedev2007,Vlahovska2007, Turitsyn2008, farutin2010analytical, abreu2013noisy}
show excellent agreement with experimental observations \cite{de1997deformation, abkarian2002tank, kantsler2005orientation, kantsler2006transition, mader2006dynamics, deschamps2009dynamics, levant2014complex}
and numerical simulations \cite{kraus1996fluid, beaucourt2004steady, noguchi2004fluid, zhao2011dynamics, yazdani2012three, Xiao2023}.

The behavior of vesicles in extensional flow, however, is less well understood.
In 2008, Kantsler \textit{et al.} studied the dynamics of strongly deflated vesicles ($\mathcal{V}<0.56$) in a planar extensional flow \cite{Kantsler2008} and discovered the existence of a critical strain rate above which vesicles undergo unbounded elongation.
\footnote{ By ``unbounded elongation'' we mean indefinite stretching within the framework of our continuum model of a fluid, incompressible membrane. At parametrically large extensions, effects such as membrane compressibility, lysis, and extreme thinning of the tether (where its radius becomes comparable to the membrane thickness $h$) become significant, requiring a different theoretical description beyond the continuum approximation.}

In this regime, the vesicle shape consists of two nearly spherical balls connected by a narrow tubular neck.
At significantly higher strain rates, vesicles exhibit a ``pearling'' instability, forming multiple bead-like structures along their length.
Extensive numerical studies of %highly elongated
{low reduced volume} vesicles in axisymmetric extensional flow were later performed in \cite{Narsimhan2014, Narsimhan2015}.

The dynamics of vesicles with moderate { reduced volume }%{elongation}
in extensional flow was investigated numerically in \cite{Zhao2013} and subsequently in \cite{Narsimhan2014}.
It was shown that vesicles transition to unbounded elongation through the development of an asymmetric instability,
where the final balls differ in size.
This transition occurs only for sufficiently deflated vesicles with $\mathcal{V}<\mathcal{V}_c \approx 0.75$,
while less deflated vesicles become more spheroidal as the strain rate increases.
Experimental studies of this asymmetric instability were reported in \cite{spjut2010trapping, Dahl2016}.
More recently, the development of feedback-based techniques for holding a vesicle near the flow focus (Stokes traps) has enabled detailed investigations of vesicle dynamics in planar extensional flows \cite{Kumar2020b, Kumar2021},
as well as relaxation after strong stretching \cite{Kumar2020}.

Further progress in this area involved studying vesicle dynamics in mixed flows \cite{deschamps2009dynamics, Lin2019},
where it was shown that the critical strain rate increases with the magnitude of the shear component.
When the strain rate varies in time \cite{kantsler2007vesicle, Lin2021}, an additional characteristic timescale emerges,
leading to three dynamic regimes for weakly deflated vesicles: pulsating, reorienting, and symmetric.

In this work, we consider the classical problem  \cite{Zhao2013, Narsimhan2014}
of a vesicle placed at the stagnation point of a uniaxial extensional flow with strain rate $\dot{\epsilon}$:
\begin{equation}
\mathbf{v}_{\mathrm{ext}} = \dot{\epsilon}\left(z\mathbf{e}_{z} - \tfrac{1}{2}\rho \mathbf{e}_{\rho}\right), \label{eq:ext_flow}
\end{equation}
where $(\rho, z)$ are cylindrical coordinates, and $\mathbf{e}_\rho, \mathbf{e}_z$ are the corresponding unit basis vectors.
Although planar flows are most commonly used in experiments \cite{Kantsler2008, Kumar2020b, Kumar2021},
many features of vesicle behavior are qualitatively similar in both cases \cite{Narsimhan2014}.

We investigate the existence and stability of steady states for different reduced volumes $\mathcal{V}$ and strain rates $\dot{\epsilon}$,
and we simulate the dynamics of the transition to unbounded elongation.
The main focus of this work is on highly elongated\footnote{  We call a shape highly elongated if it has a part whose length is substantially greater than its radius.} shapes.
For many effects, analytical asymptotic estimates can be obtained by exploiting the large aspect ratio of such vesicles.
However, from a technical perspective, the computational complexity of vesicle dynamics increases significantly with elongation.
The simulation method developed in our previous work \cite{Shishkin2026}
%\cite{link_to_archive}
%{allowed us to correct several inaccurate claims from \cite{Narsimhan2014} and to address a number of open questions.
{ enabled efficient axisymmetric simulations at high elongations, allowing us to revisit the results of \cite{Narsimhan2014}, resolve discrepancies, and address several related open questions.}

The paper is organized as follows:
Section \ref{subsec:eqns} presents the governing equations of vesicle dynamics, while Section \ref{subsec:num_scheme} describes the numerical scheme used to solve them.
In Section \ref{subsec:General_props}, we formulate a simplified model for strongly stretched vesicle shapes,
enabling us in Section \ref{susec:bifur} to determine the type of bifurcation associated with the transition to unbounded elongation,
confirmed by numerical simulations.
{Section \ref{subsec:dyno} reports simulations of vesicle dynamics in the unbounded-elongation regime.
For a strongly elongated vesicle ($\mathcal{V}\approx0.2$), we provide an analytical explanation of the observed deceleration and a direct quantitative comparison with experimental data.
For a mildly elongated vesicle ($\mathcal{V}\approx0.77$) initialized from a moderately stretched state, we show that the stationary shape remains locally stable at arbitrarily large strain rates for the given $\mathcal{V}$.}
Finally, Section 3 summarizes our findings and discusses future research directions.

%The key result of this work is the establishment of a finite critical length (contrary to the claim in \cite{Narsimhan2014})
%and the identification of the bifurcation type governing the transition of highly elongated vesicles to unbounded stretching,
%as well as the direct numerical demonstration of the metastable nature of the steady state in extensional flow.
The key finding of this work is { a quantitative characterization of vesicle metastability in extensional flow. Through direct numerical simulations and scaling analysis, we determine the extent of the stability region, the energy barriers against unbounded elongation, and their dependence on reduced volume and strain rate}.
Within this general framework, we identify the bifurcation type governing the transition of %{highly elongated}
{ low reduced volume} vesicles to unbounded stretching and establish a finite critical length, in contrast to the divergence reported in \cite{Narsimhan2014}.

\section{Governing equations}
\subsection{Dissipative Equations}\label{subsec:eqns}

In the hydrodynamic limit, lipid bilayers are treated as infinitely thin fluid films of arbitrary shape immersed in a viscous fluid.
This approach assumes that characteristic shape variations occur on length scales much larger than the bilayer thickness ($1\text{–}3\,\text{nm}$), which is valid for typical vesicle experiments \cite{Kantsler2008}.
We study the time evolution of the vesicle shape.
Hereafter, the terms \textit{membrane} and \textit{lipid bilayer} are used interchangeably.

\subsubsection{Membrane Stresses}

Macroscopically, the state of a bilayer is specified by its shape and by the surface density of lipid molecules $\displaystyle n$.
The shape of the vesicle is associated with the Helfrich bending energy \cite{Helfrich1973}:
\begin{equation}
\mathcal{F}_\kappa = \int _{S}\frac{\kappa }{2}\left(\frac{1}{R_{1}} +\frac{1}{R_{2}}\right)^{2}\mathrm{d} S, \label{eq:helfrich_energy}
\end{equation}
where integration is performed over the vesicle surface, $\kappa$ is the Helfrich bending modulus,\footnote{Typical experimental values of the Helfrich bending rigidity under normal conditions are $\kappa = 10\div25\,k_B T$, where $k_B$ and $T$ are the Boltzmann constant and the temperature \cite{Dimova2020}.} and $R_{1}, R_{2}$ are the principal radii of curvature.
\footnote{Since we do not consider processes involving changes in vesicle topology, the contribution to the free energy density proportional to the Gaussian curvature is omitted because, according to the Gauss–Bonnet theorem, this term does not change with vesicle deformation.}

Deviations of the local surface density of lipid molecules from its equilibrium value result in a surface tension $\sigma$.
Due to the weak compressibility of the membrane, the surface tension rapidly adjusts to the vesicle shape and can therefore be treated as a non-dynamic variable determined from the local inextensibility condition.
In this formulation, surface tension plays a role analogous to pressure in the Navier–Stokes equation.

The surface forces exerted on the surrounding fluid by the bilayer can be decomposed into the Helfrich force and the surface-tension force \cite{Gurin1996, Lebedev2008, zhong-can_bending_1989}:
\begin{equation}
\mathbf{f}_{\kappa } = \kappa \, \mathbf{n}\left(\frac{H^{3}}{2} -2HK+\Delta ^{\perp } H\right), \label{eq:force_hel}
\end{equation}
\begin{equation}
\mathbf{f}_{\sigma } = \nabla ^{\perp } \sigma -\mathbf{n} \,\sigma H, \label{eq:force_sigma}
\end{equation}
where $H = 1/R_{1} + 1/R_{2}$ is the mean curvature, ${K = R^{-1}_{1} R^{-1}_{2}}$ is the Gaussian curvature, $\mathbf{n}$ is the outward unit normal to the bilayer surface, and $\nabla^{\perp}$ and $\Delta^{\perp}$ denote the surface gradient and the Laplace–Beltrami operators, respectively, on the vesicle surface.

\subsubsection{Quasistatic Approximation for the Surrounding Fluid}

The inertial effects in the surrounding fluid are negligible on the characteristic length scales of vesicles ($10\text{--}100\,\mu\mathrm{m}$);
thus, the flow is governed by the Stokes equations \cite{Seifert1997}:
\begin{equation}
-\eta \Delta \mathbf{v} = \nabla p + \mathbf{F}, \label{eq:Stokes_eq}
\end{equation}
where $\mathbf{v}$ is the fluid velocity, $p$ is the pressure, $\eta$ is the dynamic viscosity, and $\mathbf{F}$ is the density of force acting on the fluid.

By balancing the rate of Helfrich energy relaxation against viscous dissipation at the characteristic vesicle scale $a$, we obtain characteristic relaxation time and flow velocity estimates:\footnote{Estimates correspond to the regime of interest where elastic stresses compete with the external forcing.}
\begin{equation}
t_\kappa = \frac{\eta a^{3}}{\kappa}, \qquad v_\kappa = \frac{\kappa}{\eta a^{2}},
\end{equation}
implying that the Reynolds number is small:
\begin{equation}
\mathrm{Re} \sim \frac{v_\kappa a \varrho}{\eta} = \frac{\kappa \varrho}{a \eta^2} \ll 1,
\end{equation}
  since the vesicle radius satisfies the condition ${a \gg {\varrho \kappa }/{\eta ^{2}} \sim 10^{-8}\,\mathrm{cm}}$\footnote{The estimate is given for water parameters.},
where $\varrho$ is the fluid density.
As the flow is slow, the pressure $p$ is determined using the incompressibility condition ${\nabla \cdot \mathbf{v} = 0}$.

Thus, the fluid velocity near the vesicle is fully determined by the instantaneous state of the bilayer,
namely, by the surface forces \eqref{eq:force_sigma}, \eqref{eq:force_hel} and by the external flow \eqref{eq:ext_flow}.

\subsubsection{Surface-Integral Representation for Stokes Flow}

When the fluids inside and outside the vesicle are identical, the membrane-induced contribution to the velocity field, $\mathbf{v}_{\mathrm{mem}}$,
can be obtained explicitly due to the linearity of the Stokes equations \eqref{eq:Stokes_eq} using the Oseen Green’s function \cite{Pozrikidis1992}:
\begin{equation}
\mathbf{v}_{\mathrm{mem}}(\mathbf{r}) = \int \mathrm{d} V_{1}\, \hat{G}(\mathbf{r} - \mathbf{r}_{1})\, \mathbf{F}_{\mathrm{mem}}(\mathbf{r}_{1}),
\end{equation}
where the subscript $1$ denotes the integration variable, and the Cartesian components of the Oseen tensor are
\begin{equation}
\hat{G}_{ik}(\mathbf{r}) = \frac{1}{8\pi \eta r}\left( \delta_{ik} + \frac{r_{i} r_{k}}{r^{2}} \right).
\end{equation}
In our case, the forces are localized on the vesicle surface, so the response is expressed as a surface integral:
\begin{equation}
\mathbf{v}_{\mathrm{mem}}(\mathbf{r}) = \int \hat{G}(\mathbf{r} - \mathbf{r}_{1})\, \mathbf{f}_{\mathrm{mem}}(\mathbf{r}_{1})\, \mathrm{d} S_{1}, \label{eq:surface_response}
\end{equation}
where $\mathbf{f}_{\mathrm{mem}}$ is the total surface force given by the sum of \eqref{eq:force_hel} and \eqref{eq:force_sigma}.
The total velocity field is then the sum of the membrane-induced velocity and the external flow \eqref{eq:ext_flow}:
\begin{equation}
\mathbf{v} = \mathbf{v}_{\mathrm{mem}} + \mathbf{v}_{\mathrm{ext}}.
\end{equation}

\subsubsection{Coupling Between Membrane Dynamics and the Surrounding Fluid}

Because the membrane is effectively impermeable and satisfies the no-slip condition,
the local velocity of the bilayer coincides with that of the fluid and is therefore continuous across the membrane.
Membrane incompressibility then requires that the surface divergence of velocity vanish:
\begin{equation}
\nabla^{\perp}\!\cdot\!\mathbf{v} = 0. \label{eq:surface_div_vel}
\end{equation}
Since, by the Stokes equation \eqref{eq:Stokes_eq}, the velocity induced by the membrane is linear in the forces applied to the fluid,
the inextensibility condition \eqref{eq:surface_div_vel} becomes a linear integral equation for the surface tension $\sigma$.
An explicit form of this equation can be obtained by expressing the surface divergence of the velocity using the Green’s function \eqref{eq:surface_response}:\footnote{The surface tension induces a contribution to the membrane stress tensor such that the surface divergence of the corresponding velocity field compensates the divergence generated by all other sources.}
\begin{multline}
\int \left( \nabla^{\perp}\hat{G}\right) \left(\nabla_{1}^{\perp}\sigma_{1} -\mathbf{n}_{1}H_{1}\sigma_{1}\right)\!\,\mathrm{d} S_{1} \ +\\
\int \!\left( \nabla^{\perp}\hat{G}\right)\!\mathbf{f}_{\kappa|1}\,\mathrm{d} S_{1} \ +
\left( \nabla^{\perp}\mathbf{v}_{\mathrm{ext}}\right) = 0, \label{eq:stension_linear_task}
\end{multline}
where the source term represents the surface velocity divergence induced by the external flow \eqref{eq:ext_flow} and the Helfrich force \eqref{eq:force_hel}.

Thus, we arrive at a closed description of vesicle-shape dynamics:
the surface tension and the surrounding flow are determined quasistatically.

\subsection{Numerical Scheme}\label{subsec:num_scheme}

\subsubsection{Parameterization of an Axisymmetric Vesicle Surface}
We consider only axisymmetric shapes; in this case, the vesicle surface can be represented by a curve on the half-plane of cylindrical coordinates $\displaystyle ( \rho, z)$, where $z$ is the axis of symmetry. This curve can be parameterized by a \textit{trajectory}  $\displaystyle \mathbf{R}( \tau ) \ =\ ( \mathcal{P}( \tau ) , \mathcal{Z}( \tau ))$, where $\tau$ is a continuous parameter. 
Note that there exists a freedom of reparameterization of the curve: one can switch to another representation $\mathbf{R}'(\tau')=\mathbf{R}(\tau(\tau'))$ with an arbitrary monotonic ``change of time'' $\tau(\tau')$. We make use of this freedom in numerical computations --- a uniform distribution of discretization nodes with respect to $\tau$ (which is convenient for algorithms) can correspond to an arbitrary spatial density of nodes, allowing refinement in “narrow” and complex regions.

Since the bilayer is advected by the flow, the dynamics of the vesicle shape are determined by the normal component of the fluid velocity:
\begin{equation}
	\partial_t \mathbf{R}\cdot \mathbf{n}=\mathbf{v}\cdot \mathbf{n},
\end{equation}
i.e., the normal part of the velocity $\partial_t \mathbf{R}$ equals the velocity of the surrounding fluid. 
The presence of reparameterization freedom implies freedom in choosing the tangential velocity, which is also used to maintain an optimal distribution of discretization nodes.

\begin{figure} \centering
	\includegraphics[width=0.8\linewidth]{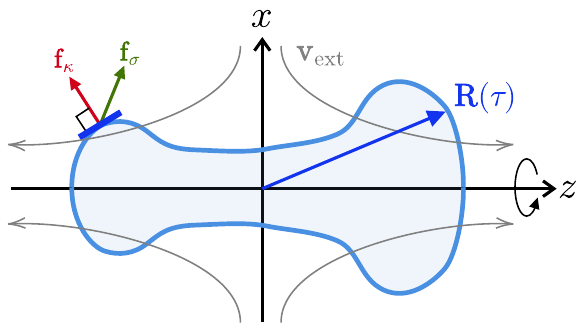}
	\caption{A schematic representation of an axially symmetric vesicle (cross-section by the $x$--$z$ plane) in an extensional flow. $\mathbf{R}(\tau)$ parameterizes the curve --- the vesicle boundary in the plane of cylindrical coordinates $(\rho,z)$. {All subsequent vesicle cross-sections in this manuscript follow this orientation.}}
	\label{fig:scheme}
\end{figure}

Since the cross-section of the vesicle by a plane containing the $z$-axis is a smooth closed curve (see Fig.~\ref{fig:scheme}\footnote{This corresponds to a symmetric continuation of the curve $\mathbf{R}(\tau)$ into the region of negative $\rho$.}), it is convenient, for numerical modeling, to use a finite Fourier series as a finite-dimensional approximation of the shape:
\begin{gather}
\begin{cases}
\mathcal{P}( \tau ) =\sum_{j=1}^N \mathcal{P}_{j} \varphi _{\rho j}\\
\mathcal{Z}( \tau ) =\sum_{j=0}^N \mathcal{Z}_{j} \varphi _{j}
\end{cases} ,\\
\varphi _{\rho j} =2\sin( \pi j\tau ) ,j\geqslant 1\ \\
\varphi _{j}      =2\cos( \pi j\tau ) ,j\geqslant 1;\ \varphi _{0} =1\ ,
\label{eq:finite_series}
\end{gather}
In view of the smoothness of the shape, the Fourier coefficients $\displaystyle ( \mathcal{P}_{j} , \mathcal{Z}_{j})$ decay exponentially with the harmonic number $\displaystyle j$.\footnote{The number of harmonics is chosen such that the relative deviations of volume and surface area from their initial values do not exceed $10^{-7}$ during the simulation.} Owing to the stiffness of the dynamical problem, we employ implicit integration schemes of the Radau type for solving the dynamical system numerically.

The expressions for the physical quantities required in the numerical modeling are given in Appendix~\ref{app:expr_quants}.
 
\subsubsection{Evaluation of Surface Integrals}

For numerical stability and robustness, the velocity $\partial_t \mathbf{R}$ is computed at more surface points $n_{\mathrm{int}}$ than the number of retained Fourier harmonics $N$.
Since the problem is axisymmetric, we employ Green’s functions analytically integrated over the azimuthal angle \cite{Pozrikidis1992}, together with their spatial derivatives.
The remaining integration in \eqref{eq:surface_response} over the parameter $\tau$ is performed numerically using local quadrature methods on a refined grid with $N_{\mathrm{int}}\gg n_{\mathrm{int}}$,
since the integral kernel contains a logarithmic singularity.

\subsubsection{Numerical Determination of Surface Tension}

Due to the smoothness of $\sigma(\tau)$, the surface tension is approximated by a truncated Fourier series \eqref{eq:finite_series}:
\begin{equation}
\sigma(\tau) = \sum_{n=0}^{N_{\sigma}} \sigma_{n}\, \varphi_{n}(\tau),
\end{equation}
which minimizes the $L_{2}$ norm of the residual error of equation \eqref{eq:stension_linear_task}:
\begin{equation}
\int \!\left[\left( \nabla^{\perp}\mathbf{v}_{\sigma}\right)[\sigma]
  + \left( \nabla^{\perp}\mathbf{v}\right)_{\text{source}}\!\right]^{2} \mathrm{d}\tau \ \rightarrow\ \min,
\end{equation}
where  $\left( \nabla^{\perp}\mathbf{v}_{\sigma}\right)$
denotes the linear integral operator mapping the surface tension $\sigma(\tau)$ to the corresponding surface velocity divergence according to \eqref{eq:stension_linear_task},
and $\left( \nabla^{\perp}\mathbf{v}\right)_{\text{source}}$ is the sum of surface divergences induced by the external flow and the Helfrich force.

The minimization condition yields a system of linear equations for the coefficients $\{\sigma_n\}$:
\begin{equation}
\sum_{n}\mathcal{O}_{kn}\sigma_{n} = -s_{k}, \label{eq:sigma_system}
\end{equation}
with
\begin{gather}
\mathcal{O}_{kn} = \int \!\left(\nabla^{\perp}\mathbf{v}_{\sigma}[\varphi_{n}]\right)
  \left(\nabla^{\perp}\mathbf{v}_{\sigma}[\varphi_{k}]\right)\mathrm{d}\tau,\\
s_{k} = \int \!\left(\nabla^{\perp}\mathbf{v}\right)_{\text{source}}
  \left(\nabla^{\perp}\mathbf{v}_{\sigma}[\varphi_{k}]\right)\mathrm{d}\tau.
\end{gather}
The system \eqref{eq:sigma_system} is solved numerically.

\section{Vesicle response to elongating flow}
\subsection{General Properties of a Stretched Vesicle}\label{subsec:General_props}
Since at equilibrium the Helfrich energy \eqref{eq:helfrich_energy} is minimized for fixed vesicle volume and surface area, elongation is associated with an additional contribution to the Helfrich energy and, correspondingly, a restoring force. In this section, we derive the scaling of this force with the vesicle length; in fact, this exposition generalizes the derivation in \cite{Kantsler2008} to the case of strong stretching.

First, we note that if one attempts to stretch a cylinder while preserving both its volume $V\sim LR^2$ and surface area $S\sim LR$, it cannot remain cylindrical --- inevitably, thinner and thicker regions form. The cylinder is constrained from elongating by its fixed volume.
The Helfrich energy of a tube of length $L$ and radius $R$ can be estimated as $\mathcal{F}_{\kappa|\mathrm{tube}}\sim \kappa L/R$; if we instead express the radius through the tube surface area $S_\mathrm{tube}\sim LR$, we have:
\begin{equation}
\mathcal{F}_\kappa(L)\sim \kappa \frac{L^2}{ S_\mathrm{tube}}. \label{eq:Helfrich_tube_energy}
\end{equation}
From this, we conclude that the lowest-energy configuration for a strongly stretched vesicle is one in which the tube contains as much area as possible, while the volume is contained in other regions. At the same time, the tube area can only decrease during stretching, since some area is required to form volume reservoirs.
We note that as elongation proceeds, the tube volume decreases, $V_{\mathrm{tube}}\sim{S_\mathrm{tube}^2}/{L}$; therefore, asymptotically, the volume reservoir takes the shape that encloses the entire volume while minimizing surface area (to minimize the energy \eqref{eq:Helfrich_tube_energy} for a fixed length), i.e. it forms a sphere of radius $R_\mathrm{ball}\to (3 V/(4\pi))^{1/3}$, and the tube area approaches its limiting value $S_\mathrm{tube}\to S - 4\pi R_\mathrm{ball}^2$. This explains, in particular, the weak dependence of the edge sphere sizes on vesicle length under strong deformation, as observed in \cite{Kumar2020a, Kumar2021}.
Thus, asymptotically, a stretched vesicle behaves like a spring of effective stiffness $\sim \kappa/S_\mathrm{tube}$. 
For example, this approach enables us to obtain estimates for the steady length $L$ of a long sedimentation-induced tether on a settling vesicle \cite{Boedec2013} by balancing the ``spring'' force and gravity: $\kappa L/S_\mathrm{tube}\sim \delta \rho g V $, where $\delta \rho$ is the density difference between the internal and external fluids, and $g$ is the gravitational acceleration.
Note that this formulation differs from the problem of pulling a thin nanotube from a large vesicle held by a micropipette \cite{Dimova2020}, where the surface tension is assumed fixed and the tube area $S_\mathrm{tube}\propto L$ increases so the Helfrich energy depends linearly on the tube length, and the force is constant.

We also note that if the vesicle is sufficiently elongated at equilibrium, there may be enough surface area to form several spherical volume reservoirs. The total area of the spheres $S_\mathrm{balls}$ containing the vesicle volume $V$ must be less than the vesicle surface area $S$:
\begin{equation}
S_\mathrm{balls} = \sum_j 4\pi R_{j}^2 < S \text{ for } V_\mathrm{balls} = V   \label{eq:condition_balls}.
\end{equation}
For example, to form two spheres of equal radius, it is necessary that $\mathcal{V}<1/\sqrt{2}$. Nevertheless, the presence of additional spheres at the same total vesicle length increases the total Helfrich energy both due to the reduction of the tube area $S_{\mathrm{tube}}$, and thus the increase in stiffness \eqref{eq:Helfrich_tube_energy}, and due to the Helfrich energy of the spheres ${\mathcal{F}_{\kappa|\mathrm{ball}}\approx 8\pi \kappa}$.
If the vesicle has a strongly elongated shape at equilibrium ($\mathcal{V}\ll1$), the volume reservoir, a sphere of radius ${R_\mathrm{ball}\sim (R_0^2 L_0)^{1/3}}$, consumes only a small fraction of the total area ${S_\mathrm{ball}\sim (R_0^2 L_0)^{2/3}\ll R_0^2L_0\sim S}$.\footnote{$R_0$ and $L_0$ are radius and length of the vesicle at equilibrium (without external forces). } That is, for such vesicles, the presence of several spheres instead of one leads to a relatively small increase in the Helfrich energy \eqref{eq:Helfrich_tube_energy}.

\subsection{Bifurcation of Unbounded Elongation in Extensional Flow}\label{susec:bifur}
Let us now consider the problem of a prolate vesicle at the focus of an extensional flow \eqref{eq:ext_flow}. If the strain rate $\dot \epsilon$ is sufficiently small, the vesicle is expected to relax to a stationary state close to equilibrium \cite{Seifert1997}.
As is known from simulations \cite{Zhao2013, Narsimhan2014, Narsimhan2015}, for the dynamics of a moderately elongated vesicle ${0.65<\mathcal{V}<\mathcal{V}_c\approx0.75}$ in a linear extensional flow \eqref{eq:ext_flow}, there exists a critical strain rate $\dot \epsilon_{c|a}$: for higher strain rates, the stationary symmetric $z\to -z$ shape becomes unstable with respect to a small asymmetric perturbation\footnote{In an extensional flow, there is always an asymmetric mode with growth rate exactly $\dot \epsilon$, corresponding to translation along the $z$ axis --- an exponential drift out of the flow focus. This translational degree of freedom does not affect the dynamics of the vesicle shape itself, and we do not consider it.}, during the development of which the vesicle also undergoes unbounded elongation over time.
However, we can restrict our consideration to symmetric $z\to - z$ configurations. In the case of highly deflated vesicles, there exists a limiting strain rate $\dot \epsilon_c$, above which the vesicle has no stationary symmetric shape and inevitably transitions to unbounded elongation \cite{Narsimhan2014}.
In \cite{Narsimhan2014}, it was found that these critical strain rates ($\dot \epsilon_c$ and $ \dot \epsilon_{c|a}$) of the flow are very close for %{highly elongated}
{ low reduced volume} vesicles. The authors also analyzed the behavior of the stationary vesicle length $L$ as a function of the strain rate $\dot \epsilon$ near the critical value and claimed that the length diverges as a power law $L\propto {(\dot \epsilon_c - \dot \epsilon)^{-\nu}}$. In the following, we demonstrate that this corresponds to a saddle-node bifurcation, in which the stationary vesicle length does not diverge at the critical point but remains finite, exhibiting a square-root behavior. We then present numerical simulation results confirming our reasoning about the type of bifurcation. In addition, we numerically investigated the spectrum of normal perturbations and found that for %highly elongated% 
{ low reduced volume} vesicles, asymmetric perturbations decay up to the point of the symmetric bifurcation $\dot \epsilon_c$, unlike the situation for moderately elongated vesicles \cite{Zhao2013, Narsimhan2014}.

\begin{figure}
	\centering
	\includegraphics[width=0.9\linewidth]{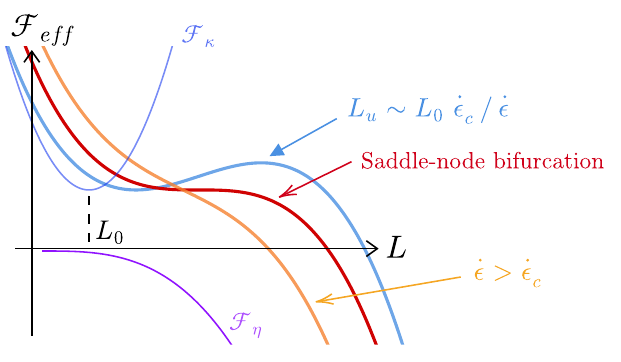}
	\caption{Qualitative picture of the bifurcation { in terms of effective free energy \eqref{eq:effective_energy}}. For strain rates below the critical value, there exists a stability region ${L<L_u\sim L_0\dot \epsilon_c/\dot \epsilon}$ beyond which the vesicle undergoes unbounded elongation. (The characteristic effective barrier height at small strain rates is $\delta \mathcal{F}_{\mathrm{eff}}\sim \kappa L_0/R_0 (\dot \epsilon_c/\dot \epsilon)^2$.) For strain rates above the critical value, the stable region disappears via a saddle-node bifurcation { at $L_u \sim L_0$}.}
	\label{fig:Bifur_schem}
\end{figure}

The starting point of the scaling analysis \cite{Kantsler2008} is the Helfrich energy for a \textit{highly elongated} configuration \eqref{eq:Helfrich_tube_energy}, from which one can estimate the force with which the vesicle resists elongation, ${F_\kappa  = -\partial_L \mathcal{F}_\kappa  \sim - \kappa {L}/{S_{\mathrm{tube}}}}$\footnote{ We would like to note that this force estimate is valid for substantial elongations $L \gtrsim L_0$, since at small extensions the volume reservoirs (balls) is only beginning to form.}, and the viscous drag force acting on the half of the vesicle due to the extensional flow, $F_\eta\sim {\eta \dot{\epsilon}L^2}/{\ln(\# L/R)}$, where $\#$ denotes a numerical factor inside the logarithm.
Comparing the length dependencies, we conclude that even at small strain rates, the previously discussed stationary state is only metastable, and for lengths exceeding the value  $L_u(\dot \epsilon) \sim \kappa/(S_{\mathrm{tube}}\eta \dot \epsilon)$ (logarithmic factors neglected), the vesicle will undergo unbounded elongation \footnote{ We note that for parametrically small strain rates, the estimate for the stability region becomes so large that effects such as membrane compressibility and lysis become significant, requiring a different theoretical description.}.
As the strain rate increases, the threshold length $L_u$ of the stable region decreases, and at some point $\dot \epsilon \sim \dot \epsilon_c \sim (\kappa/\eta)\ln(\# L_0/R_0)/(L_0^2 R_0)$ becomes of the order of equilibrium length $L_0$ and eventually disappears.
A schematic of this transition is shown in Fig.~\ref{fig:Bifur_schem} in terms of the { effective free energy:
\begin{equation}
    \mathcal{F}_\mathrm{eff} = \mathcal{F}_\kappa + \mathcal{F}_\eta \label{eq:effective_energy}
\end{equation}
sum of the Helfrich energy for the elongated configuration (with a minimum at the equilibrium length $L_0$ and quadratic asymptotics \eqref{eq:Helfrich_tube_energy} at large extensions) and the flow-stretching energy, the integrated viscous drag force along length $\mathcal{F}_\eta = \int^L F_\eta(L') \, dL'$.}
%of the integrated force, i.e., the effective free energy.
We note that the characteristic effective barrier height at small strain rates is $\delta \mathcal{F}_{eff}\sim \kappa L_0/R_0 (\dot \epsilon_c/\dot \epsilon)^2$.

Since, with increasing strain rate $\dot \epsilon$, the unstable stationary configuration at the edge of the metastable region approaches the locally stable stationary one, it is natural to expect a saddle-node type bifurcation.
Theoretically, for such a bifurcation, the stationary observables should exhibit a square-root singularity, e.g. the equilibrium elongation
${L(\dot \epsilon)/L_0-L_c/L_0\propto-\sqrt{1-\dot\epsilon/\dot\epsilon_c}}$.
Moreover, the barrier in the effective free energy also behaves as a square root ${\delta \mathcal{F}_{eff}\sim \kappa L_0/R_0 \sqrt{1-\dot \epsilon/\dot \epsilon_c}}$.
Regarding the dynamics, the growth rate of the softest normal mode\footnote{Within linear stability theory, describing the dynamics of small deviations of the shape from the stationary state $\mathbf{R}\to \mathbf{R} + \delta\mathbf{R}(t)$, normal modes are defined as perturbations that depend exponentially on time, $\delta \mathbf{R}_\lambda(t)=\delta \mathbf{R}_\lambda(0)\exp(\lambda t)$, with growth rates $\lambda$.} should vanish as a square root near the critical strain rate, ${\lambda_{\text{slowest}}\propto-\sqrt{1-\dot\epsilon/\dot\epsilon_c}}$, and correspondingly the relaxation time grows, with critical slowing down expected in the vicinity of the critical configuration.

The results of direct numerical simulations were obtained using the following procedure:
\begin{enumerate}
	\item Select a vesicle in equilibrium with a given elongation $L_0/R_0$ and corresponding reduced volume $\mathcal{V}$ (fixing the vesicle area and volume).
	\item Choose a strain rate $\dot{\epsilon}$ of the flow.
	\item Find the stationary state.
	\item In the stationary state, compute the spectrum of normal perturbations (growth rates).
	\item Repeat steps 2–4, increasing the strain rate up to the critical value.
\end{enumerate}

\begin{figure}
	\centering
	\includegraphics[width=0.9\linewidth]{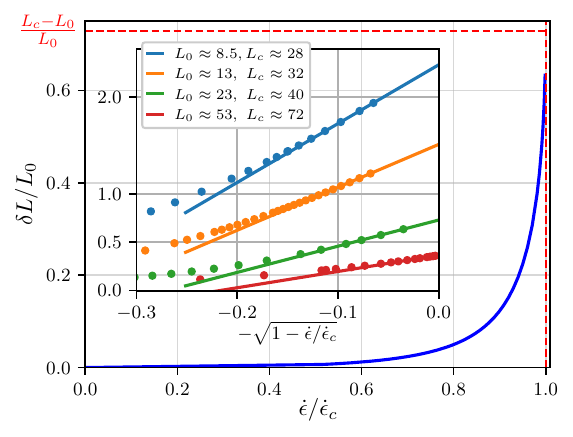}
	\caption{Stationary relative elongation as a function of strain rate $\dot \epsilon$, normalized by the critical value $\dot \epsilon_c$, for a vesicle with equilibrium elongation $L_0/R_0\approx 13 \,(\mathcal{V}\approx 0.4)$. The inset shows the same quantity in rescaled coordinates: for a saddle-node bifurcation, a square-root singularity is expected. Different colors correspond to different initial equilibrium elongations and reduced volumes $\mathcal{V} \approx {\{0.5, 0.4, 0.3, 0.2\}}$. Half-lengths $L$ are measured in equilibrium radii $R_0$. }
	\label{fig:dL_ov_eps}
\end{figure}

An example of the stationary vesicle elongation as a function of  strain rate is shown in Fig.~\ref{fig:dL_ov_eps}. A sharp increase near the critical strain rate is observed. To illustrate the square-root singularity, the inset shows the relative elongation $L/L_0-1$ for vesicles with different equilibrium elongations as a function of $(-\sqrt{1-\dot \epsilon/\dot\epsilon_c})$: for a saddle-node bifurcation, this dependence should be approximately linear near zero, which is confirmed by the plot.

\begin{figure}
	\centering
	\includegraphics[width=0.9\linewidth]{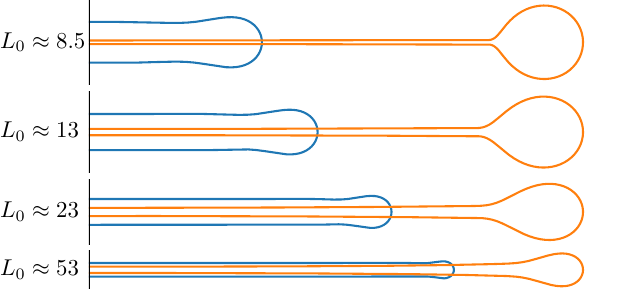}
	\caption{Comparison of equilibrium configurations (half $z>0$ shown) of vesicles and nearly critical configurations in extensional flow for different equilibrium half-lengths $L_0$ (measured in central radii $R_0$) { and corresponding $\mathcal{V}\approx \{0.5, 0.4, 0.3, 0.2\}$}.}
	\label{fig:crit_elongs_series_plot}
\end{figure}

Notably, as the reduced volume decreases (i.e., equilibrium elongation increases), the critical relative elongation decreases, as shown in Fig.~\ref{fig:dL_ov_eps}. This is further illustrated by comparing equilibrium vesicle shapes with near-critical shapes in Fig.~\ref{fig:crit_elongs_series_plot}.

\begin{figure*}
	\centering
	\subfigure[]{
		\includegraphics[height=0.25\linewidth]{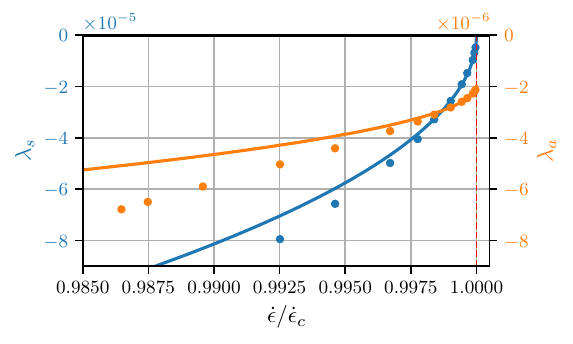} \label{fig:lambda_ov_eps_02}}
	\subfigure[]{
		\includegraphics[height=0.25\linewidth]{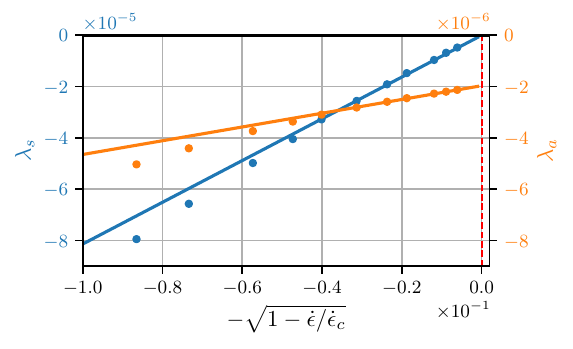} \label{fig:lambda_ov_sqrt_eps_02}}
	
	\caption{Marginal modes of the stationary vesicle with equilibrium elongation $L_0/R_0 \approx 53$ {( $\mathcal{V}\approx 0.2$)} as functions of the strain rate (\subref{fig:lambda_ov_eps_02}, \subref{fig:lambda_ov_sqrt_eps_02}). Points show simulation data; lines indicate asymptotic behavior obtained by fitting model laws close to the critical point. Growth rates $\lambda$ are measured in $t_\kappa^{-1}= \kappa/\eta R_0^{-3}$.}
	\label{fig:lambdas_02}
\end{figure*}

\begin{figure}
	\centering
	\includegraphics[height=0.55\linewidth]{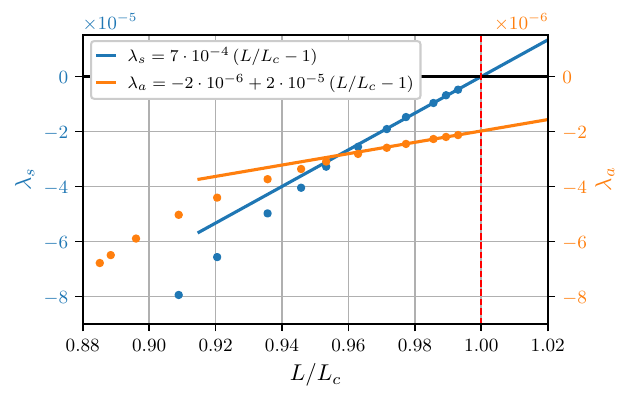} 
	\caption{Marginal modes of the stationary vesicle with equilibrium elongation $L_0/R_0 \approx 53$ as functions of the stationary vesicle length. Points show simulation data; lines indicate asymptotic behavior obtained by fitting model laws close to the critical point. Growth rates $\lambda$ are measured in $t_\kappa^{-1}= \kappa/\eta R_0^{-3}$.}
	\label{fig:lambda_ov_L_02}
\end{figure}

The results of the normal mode spectrum for stationary vesicles with equilibrium elongation $L_0/R_0\approx 53$ are shown in Fig.~\ref{fig:lambdas_02}. In Fig.~\ref{fig:lambda_ov_eps_02}, the growth rate $\lambda$ of the symmetric mode vanishes critically, while the asymmetric mode also shows a square-root singularity but remains negative.
In Fig.~\ref{fig:lambda_ov_sqrt_eps_02}, the same quantities are plotted as functions of the square root of the supercriticality, $\lambda(-\sqrt{1-\dot \epsilon/\dot \epsilon_c})$; for a saddle-node bifurcation, these dependencies should approach linearity.
Since the observed vesicle characteristics contain the same square-root singularity, plotting $\lambda(L)$ as a function of the stationary vesicle length shows a smooth dependence, as seen in Fig.~\ref{fig:lambda_ov_L_02}.
Thus, %highly elongated% 
{low reduced volume}
vesicles are stable with respect to asymmetric perturbations up to the bifurcation where the stationary state disappears.

We note that for axisymmetric geometry, the fluid velocity $\mathbf{v}$ on the vesicle surface vanishes in the stationary state, and hence there is no internal flow. That is, the contribution to the membrane force balance from the internal fluid consists only of scalar pressure, so the stationary vesicle states and bifurcation points we obtained are valid for arbitrary viscosity ratios of the vesicle interior to the surrounding fluid.\footnote{Of course, the normal mode spectra depend significantly on the properties of the internal fluid.}

\subsection{Dynamics in the Unbounded‑Elongation Regime}\label{subsec:dyno}
\subsubsection{Low Reduced Volume}
\begin{figure*}
	\centering
	\includegraphics[width=0.8\linewidth]{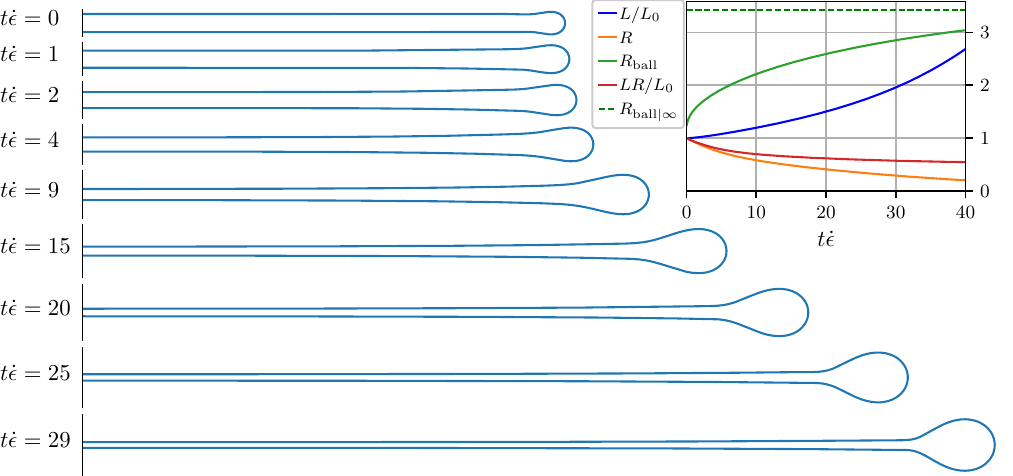}
	\caption{Dynamics of a vesicle (half shown for $z>0$) with $\mathcal{V}\approx 0.2$ in an extensional flow with a strain rate above the critical value, $\dot \epsilon /\dot \epsilon_c = 1.15$. The inset shows the time evolution of various characteristics: half-length $L$, central radius $R$, and maximum radius $R_\mathrm{ball}$. The dashed line indicates the asymptotic value of the maximum radius corresponding to a sphere containing half of the total volume, $4/3\pi R_{\mathrm{ball}|\infty}^3=V/2$.}
	\label{fig:infinite_elong_series}
\end{figure*}

We performed numerical simulations of the dynamics of vesicles stretched in an extensional flow with a rate slightly above the critical value, $\dot \epsilon \gtrsim \dot \epsilon_c$. An example of such dynamics for a vesicle with reduced volume $\mathcal{V}\approx0.2$ is shown in Fig.~\ref{fig:infinite_elong_series}.

\begin{figure}
	\centering
	\includegraphics[width=0.9\linewidth]{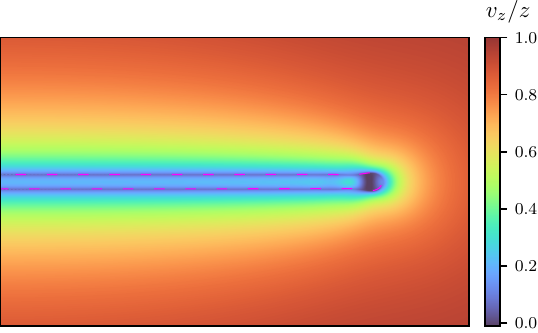}
	\caption{Reduced longitudinal velocity $v_z/z$ (in units of $\dot \epsilon$) for a stretched vesicle $\mathcal{V}\approx 0.2$ in an extensional flow. The dashed magenta line indicates the vesicle boundary.}
	\label{fig:velocity_field_Vcal=0.2_ext_flow_uz_ov_z}
\end{figure}

The main observation is that the dynamics are slowed down compared to the scaling estimate of the characteristic time based on the inverse of the strain rate, $\dot \epsilon^{-1}$.%\footnote{ We note that the obtained dynamics are consistent with the experimentally observed slowdown (elongation by a factor of $\sim 1.5$ over $t\dot{\epsilon}\sim 20$.) reported in \cite{Kantsler2008}} 
This can be explained by the effect of the incompressible membrane on the flow: Fig.~\ref{fig:velocity_field_Vcal=0.2_ext_flow_uz_ov_z} shows the reduced longitudinal velocity $v_z/z$, which would be equal to unity everywhere in the absence of the vesicle. The perturbation induced by the vesicle has a transverse scale of the order of its length. As follows from the solution of the problem of a cylindrical vesicle in an extensional flow (Appendix~\ref{app:ext_flow_around_cylinder}), the reduced velocity on the vesicle surface away from the edges is parametrically smaller than the strain rate $\dot \epsilon$:
\begin{equation}
v_z/z=\frac{\dot \epsilon}{4\ln(\tilde{L}/R)-1},\label{eq:v_z_ov_cylinder_b}
\end{equation}
where $R$ is the vesicle radius, {and $\tilde{L}$ is a parameter characterizing the external flow scale outside the vesicle, which, as we show below, is of the order of the vesicle length}.\footnote{The logarithmic slowdown was inferred from scaling analysis in \cite{Narsimhan2014}. Here, in the Appendix, we present the complete solution, including the numerical prefactor and the velocity profile both inside and outside the vesicle.}

\begin{figure}
	\centering
	\includegraphics[width=0.78\linewidth]{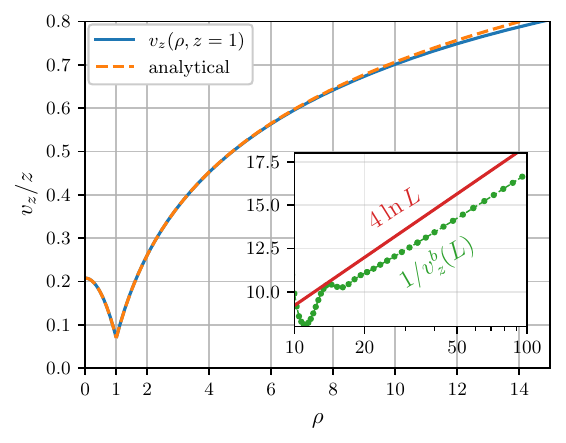}
	\caption{Profile of longitudinal velocity $v_z(\rho, z=R_0)$ for a vesicle $\mathcal{V}\approx 0.2$ ($L/R_0\approx 53$) in the extensional flow. The dashed line shows the analytical dependence \eqref{eq:v_z_ov_z_cylinder} with $\tilde{L}\approx 48$. The inset demonstrates the logarithmic decrease of the scaled velocity on the vesicle $v^b_z=v_z(\rho=R_0)/(R_0\dot\epsilon)$ with increasing vesicle length $L$ (measured in units of $R_0$). According to \eqref{eq:v_z_ov_cylinder_b}, for large $L$, $1/v^b_z$ should approach a linear asymptote.}
	\label{fig:vz_ov_z_profile}
\end{figure}

\begin{figure}
	\centering
	\includegraphics[width=0.7\linewidth]{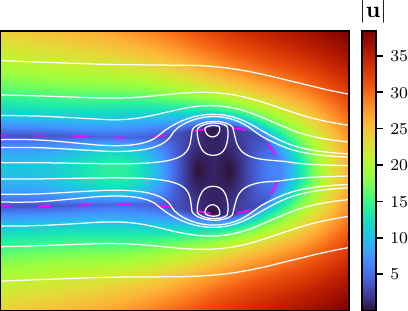}
	\caption{Flow near the vesicle edge (dashed magenta line) at the equilibrium shape, $\mathcal{V}\approx 0.2$. Color indicates the local velocity magnitude (in units of $\dot \epsilon R_0$), and white lines show instantaneous streamlines.}
	\label{fig:velocity_field_Vcal=0.2_ext_flow_appex}
\end{figure}

Comparison of the reduced longitudinal velocity profile $v_z/z$ away from the edges (at $z=R_0$) with the analytical result from Appendix~\ref{app:ext_flow_around_cylinder} is shown in Fig.~\ref{fig:vz_ov_z_profile}, showing good numerical agreement on scales smaller than the vesicle length. The inset of the same figure explicitly demonstrates the logarithmic slowdown of dynamics with increasing vesicle elongation $L/R_0$. Note that this slowdown is consistent with the well-known statement in the literature that an ellipsoidal vesicle is advected by an extensional flow in a solid-like manner (the fluid velocity inside the vesicle is zero everywhere) \cite{Narsimhan2014}: in the limit $L\to \infty$, for the inner part of the vesicle, the difference between a cylinder with edges and an ellipsoid disappears, and $v_z/z$ inside must tend to zero. Practically homogeneous flow is observed in the inner region, while near the edges the flow has a more complex structure (see Fig.~\ref{fig:velocity_field_Vcal=0.2_ext_flow_appex}).

\subsubsection{Comparison with the Experiment of Kantsler \textit{et al.} 2008}
To assess the validity of such a description and modeling of the dynamics, we performed the following comparison with experimental data. From snapshots of the dynamics of an extended vesicle in Fig.~3c of Ref.~\cite{Kantsler2008}, we extracted the vesicle length $\mathrm{Length}$ at different times $t \dot\epsilon$ (see Table~\ref{tab:kantsler_data}).
%as a function of the accumulated strain

\begin{table*}
	\centering
	\begin{tabular}{lccccccccc}
		\cmidrule(lr){2-10}
		Time, $t\dot\epsilon$ & 1.6 & 2.1 & 2.8 & 3.8 & 7.6 & 9.6 & 16.0 & 17.0 & 19.1 \\
		Length, ($\mu$m) & 66.6 & 68.2 & 69.6 & 72.6 & 77.1 & 83.1 & 98.5 & 103.0 & 115.0 \\
		\bottomrule
	\end{tabular}
	\caption{Data from the experiment of Kantsler \textit{et al.} (2008)~\cite{Kantsler2008}, Fig.~3(c).}
     \label{tab:kantsler_data}
\end{table*}

\begin{figure}
	\centering
	\includegraphics[width=0.95\linewidth]{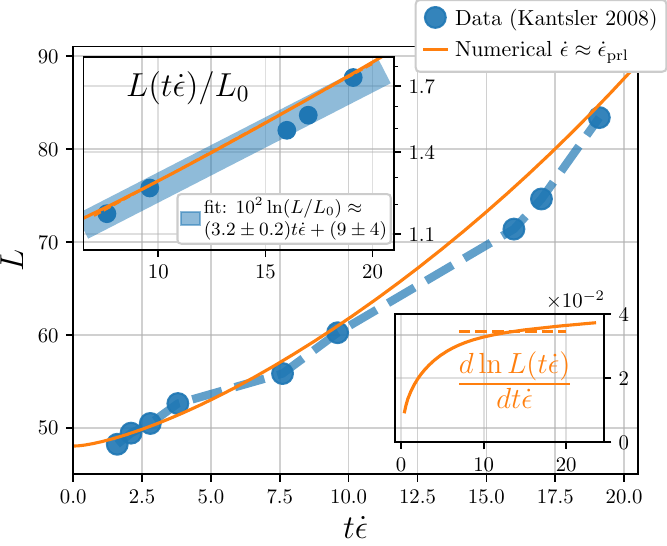}
	\caption{Comparison of the vesicle stretching dynamics (its half-length measured in units of the equilibrium radius $R_0$) between the simulation and the experimental data from Table~\ref{tab:kantsler_data}.
		The top-left inset shows the same dynamics, but with the length normalized by the equilibrium length and on a logarithmic scale, as well as the fitted straight lines (over the interval $\dot\epsilon\in[7,20]$): semi-transparent for the experimental data (width conveys the fitting uncertainty) and dashed for the model (indistinguishable over the fitting range).
		The bottom-right inset displays the derivative of the logarithm of the vesicle length with respect to $t\dot\epsilon$ for the simulated dynamics: some flattening is observed; the dashed line marks the level corresponding to the slope of the fitted straight line $\approx 3.45\times10^{-2}$.}
	\label{fig:Comparing_Kantsler}
\end{figure}

Since the exact value of the Helfrich modulus $\kappa$ is not specified in Ref.~\cite{Kantsler2008}, we performed numerical simulations for a vesicle with the same aspect ratio in equilibrium
\footnote{
Using the equilibrium diameter $D_0\approx 1.4\,\mu\mathrm{m}$ specified in Ref.~\cite{Kantsler2008}} 
$L_0/R_0\approx48$ ($\mathcal{V}\approx 0.215$), with $\dot \epsilon $ slightly below the threshold for the pearling instability for this vesicle $\dot \epsilon_{c|\mathrm{prl}}$
\footnote{At the initial moment of the dynamics simulation, ${\sigma(Z=0) R_0^2/\kappa = 3/2\cdot1.05}$},
since the caption to this dynamics in Ref.~\cite{Kantsler2008} identifies this $\dot \epsilon$ as the pearling instability boundary. The comparison is shown in Fig.~\ref{fig:Comparing_Kantsler}. To obtain at least some simple quantitative, rather than purely graphical, estimate of consistency, we exploited the observation that in a sufficiently extended configuration, when the decelerating contribution from the Helfrich force decreases, the dynamics exhibits exponential growth $\dot L\propto\dot \epsilon L$ (neglecting, at moderate times, the variation of the logarithm in Eq.~\ref{eq:v_z_ov_cylinder_b}). We indeed observe some flattening of the logarithmic derivative in the lower inset; moreover, over the interval $\dot\epsilon\in[7,20]$, when fitting a straight line to the dependence of the logarithm of the relative elongation on time, the deviation is invisible at the scale of the plot. 
We perform the same linear fit for the experimental data, for which, due to noise and the small number of points, the uncertainty is substantial, but nevertheless allows at least some averaged comparison: the fitted {slope coefficient} is $3.2 \pm 0.2$ for the data of Ref.~\cite{Kantsler2008} versus $3.45$ in the simulation. We find this satisfactorily consistent, given the extremely small number of points and also the fact that neither the exact value of $\kappa$ is known, nor was the necessary time $t \dot{\epsilon}^{-1}$ accounted for in the experiment for the transition from a randomly bent state to a cylindrical one at the initial stage of the dynamics.
\footnote{We emphasize once again that we do not claim that the dynamics is already exponential over this interval, since the Helfrich force still substantially slows the dynamics: without its account, the estimate of the logarithmic derivative from Eq.~\ref{eq:v_z_ov_cylinder_b} would give a slope $\approx 4.5$; we merely use some simple way of averaging over the data via fitting to a low-parameter model---a straight line.}

\subsubsection{Moderate Reduced Volume}
As is known from \cite{Zhao2013, Narsimhan2014}, the critical  strain rate exists only for sufficiently elongated vesicles, $\mathcal{V}< \mathcal{V}_c\approx 0.75$. For larger $\mathcal{V}$, a symmetric ($z\to-z$) stationary state is locally stable for any $\dot \epsilon$.

\begin{figure}
	\centering
	\includegraphics[width=0.9\linewidth]{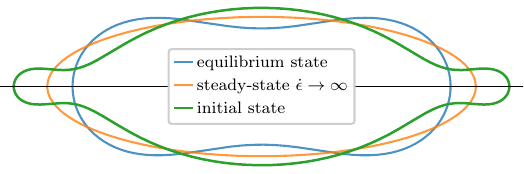}
	\caption{Equilibrium configuration, stationary state in a high-amplitude extensional flow, and the initial elongated configuration used for simulating further elongation of a vesicle with $\mathcal{V}=0.77$.}
	\label{fig:Comparing_initial_with_states}
\end{figure}

To demonstrate the metastable character of any equilibrium in an extensional flow, we performed numerical simulations of symmetric ($z\to-z$) dynamics for a slightly elongated vesicle with reduced volume ${\mathcal{V}=0.77}$. To reach unbounded elongation, such a vesicle must first be prepared in a slightly stretched state, with a transverse scale somewhat smaller than the vesicle size due to volume conservation. The initial vesicle state, starting from which it can switch to unbounded elongation in the extensional flow  under a sufficiently high strain rate is shown in Fig.~\ref{fig:Comparing_initial_with_states}, compared to the equilibrium and the limiting $\dot\epsilon \to \infty$ stationary state.

\begin{figure}
	\centering
	\includegraphics[width=0.7\linewidth]{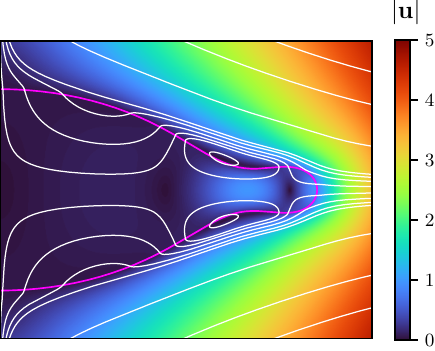}
	\caption{Flow deformation (omitting the Helfrich force contribution) around the vesicle $\mathcal{V}=0.77$ (magenta line) in the prepared state. Color indicates the local velocity magnitude (in units of $\dot \epsilon R_0$), and white lines show instantaneous streamlines.}
	\label{fig:velocity_field_Vcal=0.77_ext_flow}
\end{figure}

\begin{figure}
	\centering
	\includegraphics[width=0.7\linewidth]{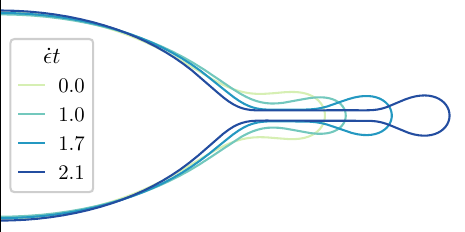}
	\caption{Dynamics of a slightly pre-stretched vesicle $\mathcal{V}\approx 0.77$ in an extensional flow with rate $\dot \epsilon \eta/(\kappa R_0^3)=7$, where $R_0$ is the central radius of the vesicle at equilibrium ($R_0\approx (V/22)^{1/3}$ for $\mathcal{V}=0.77$).}
	\label{fig:Elongating_Vcal=0.77_series_plot}
\end{figure}

Fig.~\ref{fig:velocity_field_Vcal=0.77_ext_flow} illustrates the flow around the incompressible vesicle in the prepared state ($\mathcal{V}=0.77$) for visualizing the overcoming the geometric barrier (neglecting the Helfrich force contribution, as the focus is on the possibility of transition to unbounded elongation at high strain rate). The beginning stage of unbounded elongation dynamics at $\dot \epsilon = 7\kappa/(\eta R_0^3)$ for this initial state is shown in Fig.~\ref{fig:Elongating_Vcal=0.77_series_plot}. We conclude that transition to unbounded elongation, surmounting geometric (and energetic) barriers, is possible for $z\to-z$ symmetric configurations, with the limiting shape being a large central sphere and two small spheres at the edges.

Note that transition to unbounded elongation is also possible from an asymmetric ($z\to - z$) initial state, since for any reduced volume $\mathcal{V}<1$ there exist strongly elongated asymmetric configurations \cite{Zhao2013}, in the form of two spheres of different radii at the edges connected by a tube. Qualitatively, the energy barrier is expected to be smaller than for the symmetric case, as only two spheres form instead of three. Moreover, according to condition \eqref{eq:condition_balls} for asymmetric configurations, the size ratio of the final spheres is smaller than in the symmetric case: for $\mathcal{V}=0.75$, an elongated configuration with the limiting edge-sphere radius ratio $\approx 2/3$ is possible, compared to the edge-to-central radius ratio $\approx 2/5$ for the symmetric configuration.

Furthermore, near the critical reduced volume $\mathcal{V}_c$, the energy barrier for asymmetric shape perturbations is parametrically small (see Appendix~\ref{app:critical_asym_barier}), and at sufficiently high strain rate it scales as $\delta \mathcal{F}_{\mathrm{eff}}\sim \kappa (\mathcal{V}-\mathcal{V}_c)$, i.e. in the relatively narrow range of reduced volumes $(\mathcal{V}-\mathcal{V}_c)\sim T/\kappa$, a thermally activated transition to unbounded elongation is possible.

%\FloatBarrier
\section*{Conclusion}
In this paper, we study the shape dynamics of vesicles in an axisymmetric extensional flow. Refining the analysis of the Helfrich elastic energy and the viscous stress from the flow for elongated configurations presented in \cite{Kantsler2008}, we establish the metastable nature of any stationary vesicle state in an extensional flow and estimate the extent of the stability region.

For vesicles with small reduced volume $\mathcal{V}$ (corresponding to large ratios of semi-axes at equilibrium), this analysis allows us to determine the type of bifurcation at which a stationary state disappears at a critical flow strain rate, leading the vesicle to undergo unbounded elongation. The critical elongation of the vesicle is finite, in contrast to the power-law divergence reported in \cite{Narsimhan2014}.
The obtained critical scaling is verified through direct numerical simulations (see Fig. \ref{fig:dL_ov_eps}). We find that the magnitude of the critical relative elongation decreases as the aspect ratio of the vesicle at equilibrium increases (see Fig. \ref{fig:crit_elongs_series_plot}).
Our analysis of the spectrum of normal perturbations for strongly elongated vesicles indicates that asymmetric ($z\to-z$) modes remain stable up to the bifurcation point unlike the case of vesicles with moderate equilibrium elongation \cite{Zhao2013, Narsimhan2014, Dimova2020}, where at some strain rate the stationary state becomes unstable to asymmetric perturbations.

%Furthermore, we numerically simulate the beginning stage of the dynamics for an extension rate above critical, where the vesicle eventually undergoes unbounded elongation (see Fig. \ref{fig:infinite_elong_series}). To explain the observed deceleration of these dynamics, we obtain an analytical solution of the problem of an incompressible cylindrical vesicle in an extensional flow in Appendix \ref{app:ext_flow_around_cylinder}; the resulting logarithmic slowdown is confirmed by numerical analysis in Fig. \ref{fig:vz_ov_z_profile}.

{
Furthermore, we numerically simulate the beginning stage of the dynamics for an extension rate above critical, where the vesicle eventually undergoes unbounded elongation (see Fig.~\ref{fig:infinite_elong_series}). To explain the observed deceleration of the dynamics, we obtain an analytical solution of the problem of an incompressible cylindrical vesicle in an extensional flow in Appendix~\ref{app:ext_flow_around_cylinder}; the resulting logarithmic slowdown is confirmed by numerical analysis in Fig.~\ref{fig:vz_ov_z_profile}.\footnote{We note that such deceleration of vesicle shape dynamics makes experimental observation more challenging, as small displacements from the stagnation point along the $z$-axis cause the vesicle as a whole to drift away exponentially at rate $\dot \epsilon$. Although this translational motion is decoupled from the shape dynamics, the relatively rapid escape of the vesicle from the flow focus requires feedback systems that adjust the focus of the extensional flow in response to vesicle displacement, such as the so-called Stokes trap for planar flow \cite{shenoy2016stokes, Kumar2020}.} Moreover, we performed a direct quantitative comparison of this dynamics with experimental data from ~\cite{Kantsler2008} (see Fig.~\ref{fig:Comparing_Kantsler}), finding satisfactory agreement in the exponential growth rates between simulation and experiment, which validates the predictive capability of the model for highly elongated vesicles. }

For vesicles with reduced volume ${\mathcal{V} > \mathcal{V}_c\approx 0.75}$, it is known from the literature \cite{Zhao2013,Narsimhan2014, Dimova2020} that there is no critical strain rate: for any value, a stationary $z\to-z$ symmetric vesicle state is locally stable.
Through direct numerical simulations (Fig. \ref{fig:Elongating_Vcal=0.77_series_plot}) of a vesicle with $\mathcal{V}\approx0.77$ initialized in a moderately elongated state, we demonstrate the metastable nature of the equilibrium even with respect to $z\to-z$ symmetric deformations.
Using a phenomenological model for the free energy as a function of vesicle asymmetry near the stationary state, we show in Appendix \ref{app:critical_asym_barier} that in the vicinity of $\displaystyle \mathcal{V}_{c}$, the energy barrier to unbounded elongation via asymmetric perturbations is parametrically small, so thermally activated transitions across this barrier are expected experimentally within a certain range of strain rates.

A natural extension of the study would be to apply the same physical and numerical framework to investigate vesicle behavior in electromagnetic fields, since experimental manipulation of biological particles using optical tweezers offers several advantages, and to generalize the scheme to arbitrary perturbations without assuming axisymmetry. This would allow not only quantitative comparison with planar extensional flow experiments \cite{Dahl2016,Kumar2020a, Kumar2020b, Kumar2021}, but also the consideration of thermal fluctuations. We anticipate that this work will be useful for understanding the elastic properties of elongated vesicles, which is important for potential biomedical applications.
 \section*{Acknowledgment}

During this work, we benefited greatly from many profound and fruitful discussions with Professor V.V. Lebedev, which are gratefully acknowledged. 
The investigation of stationary vesicle shapes in the extensional flow was supported by the Russian Science Foundation (Grant No. 23-72-30006). The numerical simulation of the beginning stage of unbounded elongation dynamics was supported by the Ministry of Science and Higher Education of the Russian Federation (state assignment no. FFWR-2024-0014) of Landau Institute for Theoretical Physics of the RAS. 
M.A.Sh. gratefully acknowledges the Foundation for the Advancement of Theoretical Physics and Mathematics ``BASIS'' for support which enabled the analytical results presented in Appendixes.

\section*{Competing Interests}
The authors declare that they have no competing interests.
%\printbibliography
%\bibliography{All_bib}
%\clearpage
\appendix
\numberwithin{equation}{section}
\section{Physical quantities expressions}\label{app:expr_quants}
Below we present expressions for the physical quantities required to determine the velocity in a general reparameterization of an axisymmetric surface $\displaystyle \mathbf{R}( \tau )$.
The first fundamental form of the surface is determined by the metric in cylindrical coordinates $\displaystyle ( \rho ,z,\varphi )$:
\begin{equation*}
ds^{2} \ =\ dz^{2} +d\rho ^{2} +\rho ^{2} d\varphi ^{2} \ =\ {\left( \mathcal{Z}_{\tau }^{2} +\mathcal{P}_{\tau }^{2}\right)} d\tau ^{2} +\mathcal{P}^{2} d\varphi ^{2},
\end{equation*}
where $\mathcal{P}_\tau$ and $\mathcal{Z}_\tau$ denote the corresponding first derivatives with respect to the parameter $\tau$.
The area element is then given by
\begin{equation*}
\mathrm{d} S\ =\ \left( \mathcal{Z}_{\tau }^{2} +\mathcal{P}_{\tau }^{2}\right)^{1/2}\mathcal{P}\ d\tau d\varphi .
\end{equation*}
In particular, the surface area and the vesicle volume are
\begin{equation*}
S\ =\ 2\pi \int _{0}^{1} \left( \mathcal{Z}_{\tau }^{2} +\mathcal{P}_{\tau }^{2}\right)^{1/2}\mathcal{P}\ d\tau ,\ V\ =\ \pi \int _{0}^{1} \mathcal{P}^{2} \mathcal{Z}_{\tau } d\tau .
\end{equation*}
The surface gradient operator is
\begin{equation*}
\nabla ^{\perp } \ =\ \mathbf{e}_\tau\frac{1}{\left( \mathcal{Z}_{\tau }^{2} +\mathcal{P}_{\tau }^{2}\right)^{1/2}} \partial _{\tau } \ +\ \mathbf{e}_{\varphi }\frac{1}{\mathcal{P}} \partial _{\varphi } ,
\end{equation*}
where $\mathbf{e}_\varphi$ is the unit basis vector corresponding to the azimuthal angle $\varphi$, and the tangent vector in the $( \rho, z)$-plane is
\begin{equation*}
\mathbf{e}_\tau \ =\frac{\begin{pmatrix}
	\mathcal{P}_{\tau }, & \mathcal{Z}_{\tau }
	\end{pmatrix}}{\left( \mathcal{Z}_{\tau }^{2} +\mathcal{P}_{\tau }^{2}\right)^{1/2}}.
\end{equation*}
The unit normal vector is defined as
\begin{equation*}
\mathbf{n} \ =\ \frac{\begin{pmatrix}
	-\mathcal{Z}_{\tau }, & \mathcal{P}_{\tau }
	\end{pmatrix}}{\left( \mathcal{Z}_{\tau }^{2} +\mathcal{P}_{\tau }^{2}\right)^{1/2}}.
\end{equation*}
The expressions for the principal curvatures are then:
\begin{gather*}
H_{1} =\frac{\mathcal{Z}_{\tau } \mathcal{P}_{\tau \tau } -\mathcal{P}_{\tau } \mathcal{Z}_{\tau \tau }}{\left( \mathcal{Z}_{\tau }^{2} +\mathcal{P}_{\tau }^{2}\right)^{3/2}} ,\ H_{2} =-\frac{\mathcal{Z}_{\tau }}{\mathcal{P}\left( \mathcal{Z}_{\tau }^{2} +\mathcal{P}_{\tau }^{2}\right)^{1/2}},\\
H = H_1+H_2, K = H_1 H_2.
\end{gather*}
The Laplace-Beltrami operator is given by
\begin{equation*}
\Delta^\perp=\frac{1}{\left( \mathcal{Z}_{\tau }^{2} +\mathcal{P}_{\tau }^{2}\right)^{1/2}\mathcal{P}}\left( \partial _{\tau }\frac{\mathcal{P}}{\left( \mathcal{Z}_{\tau }^{2} +\mathcal{P}_{\tau }^{2}\right)^{1/2}} \partial _{\tau }\right) +\frac{1}{\mathcal{P}^{2}} \partial _{\varphi }^{2}. \end{equation*}
These relations allow the derivation of an explicit form of the Helfrich force in an arbitrary parametrization in terms of the functions $\mathcal{P}(\tau), \mathcal{Z}(\tau)$ and their derivatives up to fourth order.

We also present the necessary expressions for the surface divergence of the velocity in cylindrical coordinates:
\begin{equation}
    \nabla ^{\perp }\cdot \mathbf{v} =\frac{\mathcal{P}_{\tau }}{\left( \mathcal{Z}_{\tau }^{2} +\mathcal{P}_{\tau }^{2}\right)} \partial _{\tau } v_{\rho } +\frac{\mathcal{Z}_{\tau }}{\left( \mathcal{Z}_{\tau }^{2} +\mathcal{P}_{\tau }^{2}\right)} \partial _{\tau } v_{z} \ +\ \frac{{v_{\rho }}}{\mathcal{P}}.\label{eq:surface_div_v}
\end{equation}

For the linear extensional flow \eqref{eq:ext_flow}, using \eqref{eq:surface_div_v} we obtain
$$
\nabla ^{\perp }\mathbf{v}_{\mathrm{ext}}= \dot\epsilon \frac{-2\mathcal{P}_{\tau }^{2} +\mathcal{Z}_{\tau }^{2}}{2\left( \mathcal{Z}_{\tau }^{2} +\mathcal{P}_{\tau }^{2}\right)}.
$$
 \section{Viscous flow around an incompressible cylinder}\label{app:ext_flow_around_cylinder}
In this section, we present a solution to the canonical problem of flow near a long cylindrical vesicle of unit radius in a uniaxial extensional flow with unit strain rate. We employ the stream function formalism $\psi(\rho, z)$ for axisymmetric flow \cite{happel_low_1983}: 
\begin{equation}
v_{z} = -\frac{1}{\rho} \partial_{\rho} \psi, \quad v_{\rho} = \frac{1}{\rho} \partial_{z} \psi.
\end{equation}
The stream function of the extensional flow is given by $\psi_{\rm ext} = -\frac{\rho^2}{2} z$. The Stokes equation \eqref{eq:Stokes_eq} reduces to
\begin{equation}
E^4 \psi = 0, \quad E^2 = \rho \partial_\rho \frac{1}{\rho} \partial_\rho + \partial_z^2.
\label{eq:psi_eq}
\end{equation}

We aim to describe the flow far from the vesicle edges, and thus employ the ansatz $\psi = z f(\rho)$. As a consequence of equation \eqref{eq:psi_eq}, the function $f$ is locally a linear combination of $\{\rho^2, 1, \rho^4, \rho^2 \ln(\tilde{L}/\rho)\}$, where $\tilde{L}$ is a parameter characterizing the outer flow scale. From physical considerations, $\tilde{L}$ is of the order of the vesicle length\footnote{This parameter cannot be determined from the solution far from the vesicle edges, since there are no two linearly independent flows with this symmetry that decay at infinity in $\rho$.}. Due to the smoothness of the velocity field inside the vesicle, the solution is a linear combination of $\{\rho^2, \rho^4\}$. Outside the vesicle, the correction to $\psi_{\rm ext}$ due to the presence of the vesicle should not grow rapidly, so $f(\rho)$ outside can be considered a linear combination of $\{1, \rho^2 \ln(\tilde{L}/\rho)\}$.

On the vesicle surface, the following conditions must be satisfied:
\begin{enumerate}
\item Continuity of the stream function.
\item Continuity of the fluid velocity.
\item Vanishing surface divergence of the velocity, $\nabla^\perp \cdot \mathbf{v} = \partial_z v_z + v_\rho/\rho$.
\item Tangential and normal force balance due to the presence of some surface tension $\sigma(z)$.
\end{enumerate}

The unknown surface tension can be eliminated from the force balance conditions:
\begin{multline}
\begin{array}{l}
\hat{\sigma}_{\rho \rho|\mathrm{in}} - \hat{\sigma}_{rr|\mathrm{out}} = -\sigma \\
\hat{\sigma}_{\rho z|\mathrm{in}} - \hat{\sigma}_{\rho z|\mathrm{out}} = \partial_z \sigma
\end{array} \implies \\
\partial_z \hat{\sigma}_{\rho\rho|\mathrm{in}} + \hat{\sigma}_{\rho z|\mathrm{in}} = \partial_z \hat{\sigma}_{\rho \rho|\mathrm{out}} + \hat{\sigma}_{\rho z|\mathrm{out}},
\end{multline}
where the stresses are defined as $\hat{\sigma}_{\rho\rho} = -p + 2\eta \partial_\rho v_\rho$, $\hat{\sigma}_{z\rho} = \eta(\partial_\rho v_z + \partial_z v_\rho)$, and the pressure gradient is $\partial_z p = -\eta/\rho \, \partial_\rho E^2 \psi$.

Solving the resulting system of linear equations for the coefficients in the basis expansion yields the following expressions for the stream function:
\begin{multline}
f_{\rm in} = \frac{1/2}{4 \ln \tilde{L} - 1} \left(-3\rho^2 + \rho^4 \right),\\
f_{\rm out} = -\frac{1}{2} \rho^2 + \frac{-3 + 4 \rho^2 \ln(\tilde{L}/\rho)}{2(4 \ln \tilde{L} - 1)},
\end{multline}
from which $v_\rho = f/\rho$, while $v_z$ takes the form
\begin{equation}
v_z^{\rm in}/z = \frac{3 - 2 \rho^2}{4 \ln \tilde{L} - 1}, \quad
v_z^{\rm out}/z = 1 - 2 \frac{2 \ln (\tilde{L}/\rho) - 1}{4 \ln \tilde{L} - 1}.
\label{eq:v_z_ov_z_cylinder}
\end{equation}

{ We note that this result clearly implies that the cylindrical part of the vesicle cannot be stationary due to the Helfrich force contribution, as it merely adds a constant contribution to the pressure difference on an ideal cylinder \eqref{eq:force_hel} without affecting the velocity. In stationary states at $\dot{\epsilon}<\dot{\epsilon}_c$, the balance is instead achieved because the vesicle is no longer cylindrical: there is a modulation of the radius on a scale of order the vesicle length, since the surface derivative of the Helfrich force is no longer zero.}

\section{Estimation of energy barrier near  $\mathcal{V}_c$}\label{app:critical_asym_barier}
As is known from \cite{Narsimhan2014}, in the limit of a large strain rate, $\dot \epsilon t_\kappa \gg 1$, the stationary shape of the vesicle is a spheroid. In this case, the growth rate $\lambda$ of the marginally asymmetric perturbation $z\to -z$ crosses zero regularly at the reduced volume $\mathcal{V}_c \approx 0.75$, i.e. near the critical point $\lambda \propto (\mathcal{V}_c - \mathcal{V})$. Then, in the same vicinity, one can write an effective free energy associated with the asymmetric deformation as a function of the dimensionless asymmetry $s$:
\begin{equation}
\mathcal{F}_{\mathrm{eff}|\mathrm{ext}} = \dot{\epsilon}\eta R^3 \big(\alpha_{\mathrm{ext}}(\mathcal{V}-\mathcal{V}_c) s^2 - \beta_{\mathrm{ext}}s^4 \big),
\end{equation}
where $\alpha_{\mathrm{ext}}$ and $\beta_{\mathrm{ext}}$ are some positive numerical coefficients of order unity, and $R$ is a characteristic size of the vesicle.

If we take into account a small Helfrich constant, $\kappa \ll \dot \epsilon \eta R^3$, the stationary shape slightly deviates from the spheroidal form, so the coefficients $\alpha_{\mathrm{ext}}(\mathcal{V}-\mathcal{V}_c), \beta_{\mathrm{ext}}$ are slightly modified by an amount of order $\kappa / (\dot \epsilon \eta R^3)$. In addition, the contribution to the free energy from the Helfrich bending energy $\sim \kappa$ must also be considered.  
Thus, accounting for the small ratio $\kappa/(\dot \epsilon \eta R^3)$, the full effective free energy takes the form:
\begin{equation}
\mathcal{F}_{\mathrm{eff}} = \dot{\epsilon}\eta R^3 \big(\alpha_{\mathrm{ext}}(\mathcal{V}-\mathcal{V}_c) s^2 - \beta_{\mathrm{ext}}s^4 \big) + \kappa\alpha_\kappa s^2,\label{eq:F_asym_eff}
\end{equation}
where $\alpha_\kappa$ is a positive number of order unity.

From this, one can immediately reproduce the critical behavior of $\dot \epsilon$ in the region $\mathcal{V} < \mathcal{V}_c$:
\begin{equation}
	\dot \epsilon_c = \frac{\kappa}{\eta R^3} \frac{\alpha_\kappa}{\alpha_\mathrm{ext}} \frac{1}{\mathcal{V}_c - \mathcal{V}}.\label{eq:power_low}
\end{equation}
To our knowledge, the divergence of the critical rate is reported in the literature, but the power-law behavior \eqref{eq:power_low} has not been established \cite{Zhao2013, Narsimhan2014, Dimova2020}.  
Note that in experiments, even at a strain rate below the critical value, a transition to unbounded stretching can be observed due to thermal barrier crossing, whose magnitude can be estimated from \eqref{eq:F_asym_eff}:
\begin{multline}
{s_c^2 \sim (\mathcal{V}_c - \mathcal{V}) \frac{\dot \epsilon_c - \dot \epsilon}{\dot \epsilon }},\\
\delta\mathcal{F}_\mathrm{eff} \sim \kappa (\mathcal{V}_c - \mathcal{V}) \frac{(1 - \dot\epsilon/\dot\epsilon_c)^2}{\dot\epsilon/\dot\epsilon_c}.
\end{multline}
Comparing this value with the thermal energy, we find that in the region $\mathcal{V}_c - \mathcal{V} \gg k_B T / \kappa$, the strain rate $\dot \epsilon_T$, at which frequent thermal transitions to unbounded stretching can be observed, is close to $\dot \epsilon_c$:
\begin{equation}
	1 - \dot\epsilon_T/\dot\epsilon_c \sim \sqrt{ (k_B T / \kappa) / (\mathcal{V}_c - \mathcal{V}) },
\end{equation}
whereas in the narrow region $\mathcal{V}_c - \mathcal{V} \lesssim k_B T / \kappa$ of reduced volumes, the value of $\dot \epsilon_T$ remains of order
\begin{equation}
	\dot\epsilon_T \sim \frac{\kappa}{\eta R^3} \frac{\kappa}{k_B T},
\end{equation}
and does not diverge at the critical point.

Now consider the region $\mathcal{V} > \mathcal{V}_c$, where the stationary state is locally stable for any $\dot \epsilon$. In this case, the critical asymmetry amplitude $s_c$ is
\begin{equation}
	s_c^2 \sim \frac{\kappa \alpha_\kappa}{\dot \epsilon \eta R^3} + \alpha_\mathrm{ext} (\mathcal{V} - \mathcal{V}_c),
\end{equation}
and the corresponding energy barrier is
\begin{equation}
\delta\mathcal{F}_\mathrm{eff} \sim \dot \epsilon \eta R^3 \left( \frac{\kappa \alpha_\kappa}{\dot \epsilon \eta R^3} + \alpha_\mathrm{ext} (\mathcal{V} - \mathcal{V}_c) \right)^2,
\end{equation}
which attains a minimum at the strain rate $\dot \epsilon_{\mathrm{opt}} \sim \kappa / (\eta R^3) / (\mathcal{V} - \mathcal{V}_c)$ with magnitude
\begin{equation}
	\delta\mathcal{F}_{\mathrm{eff}|\min} \sim \kappa (\mathcal{V} - \mathcal{V}_c),
\end{equation}
while the critical asymmetry amplitude is parametrically small, $s_c \sim \sqrt{\mathcal{V} - \mathcal{V}_c}$.  
For larger strain rates, the critical deformation remains of the same order, but the barrier increases, $\delta\mathcal{F}_{\mathrm{eff}}(\dot\epsilon \gtrsim \dot \epsilon_{\mathrm{opt}}) \sim \dot \epsilon \eta R^3 (\mathcal{V} - \mathcal{V}_c)^2$. For smaller rates in the intermediate region $\kappa / (\eta R^3) \ll \dot{\epsilon} \lesssim \dot \epsilon_{\mathrm{opt}}$, the barrier is also higher, $\delta\mathcal{F}_\mathrm{eff} \sim \kappa^2 / (\dot\epsilon \eta R^3)$.  
The region of very small strain rates, $\dot \epsilon \lesssim \kappa / (\eta R^3)$, cannot be described using this method, since the critical elongation becomes parametrically large (for both symmetric and asymmetric perturbations).  

A clear implication for experiments is that in a sufficiently narrow region $\mathcal{V} - \mathcal{V}_c \lesssim k_B T / \kappa$, within some range of strain rates 
\begin{equation}
\frac{\kappa^2}{k_B T} \lesssim \dot \epsilon (\eta R^3) \lesssim \frac{k_B T}{(\mathcal{V} - \mathcal{V}_c)^2},    
\end{equation}
transitions to unbounded asymmetric stretching via thermal barrier crossing should be observable.

\bibliography{All_bib}
\end{document}